\documentclass{article}%
\usepackage{amsmath}%
\usepackage{amsfonts}%
\usepackage{amssymb}%
\usepackage{graphicx}
\newtheorem{theorem}{Theorem}
\newtheorem{acknowledgement}[theorem]{Acknowledgement}

\begin{document}

\title{Dynamic Elastic Properties of  Human Bronchial Airway Tissues}
\author{J.-Y. Wang$^{a}$, P. Mesquida$^{a,b}$, P. Pallai$^{c}$, C. J. Corrigan$^{c}$, T. H. Lee$^{c}$\\
\small{jauyi.wang@kcl.ac.uk}, \small{patrick.mesquida@kcl.ac.uk}\\
$^a$\small{Division of Engineering, King's College London, Strand, London WC2R 2LS, UK}\\
$^b$\small{Department of Physics, King's College London, Strand, London WC2R 2LS, UK}\\
$^c$\small{Asthma, Allergy and Lung Biology Division, MRC-Asthma UK Centre in Allergic}\\ \small{Mechanisms of Asthma, King's College London, Guy's Hospital, London SE1 9RT, UK}
}
\date{November 20, 2011}
\maketitle

\begin{abstract}
Young's Modulus and dynamic force moduli were measured on human bronchial airway tissues by compression. A simple and low-cost system for measuring the tensile-strengh of soft bio-materials has been built for this study. The force-distance measurements were undertaken on the dissected bronchial airway walls, cartilages and mucosa from the surgery-removed lungs donated by lung cancer patients with COPD. Young's modulus is estimated from the initial slope of unloading force-displacement curve and the dynamic force moduli (storage and loss) are measured at low frequency (from 3 to 45\! Hz). All the samples were preserved in the PBS solution at room temperature and the measurements were perfomed within 4 hours after surgery. Young's modulus of the human bronchial airway walls are fond ranged between 0.17 and 1.65\! MPa, ranged between 0.25 to 1.96\! MPa for cartilages, and between 0.02 to 0.28\! MPa for mucosa. The storage modulus are found varying 0.10\! MPa with frequency while the loss modulus are found increasing from 0.08 to 0.35\! MPa with frequency. The frequency-dependent dynamic force moduli are also compared with different strain rates.
\end{abstract}

\section{Introduction}

Ariway diseases such as asthma are related to the airway wall elasticity. The elastical properties of lung tissue are important to determine the mechanical behoviour of lungs ~\cite{Zin_Rev}. Over the past few decades, dynamic measurments on the respiratory gas pressure, volume, and flow have been applied on clinical diagonis for understanding the lung mechanics such as spirometry and forced-oscillation technique~\cite{FOT_spiro}. The lung resistance $R_r$ and elastance $E_r$ are determined by the equation of motion $P(t)=E_rV(T)+R_r\dot{V}(t)+P_0$, where $P(t)$ is the time-dependent applied pressure, $V$ is the volume, and $\dot{V}$ is the flow, and $P_0$ is the pressure corresponding to transpulmonary pressure at end expiration~\cite{RE_equation}. 

The dynamical elasticity measurements on airway walls can lead to a better understanding of respiratory ventilation mechanisms and also provide useful information and reference values for clinical tools, such as ultrasound and MRI elastography [Ophir1997,Miller2007]. %\cite

Stiffness measurements on human and animal trachea rings have been studied in the past two decades [Bert1992,1997,2005]. However, the elastic properties of airway walls from bronchi are little documented. In this study, a simple and low-cost system has been developed for measuring the elastic properties of soft bio-materials. Force-distance measurements on bronchial wall segments from lung cancer patients with COPD have been performed.

\section{Basic Principles}
\paragraph{Determining Young's Modulus from Force-Distance Curve}\hspace{5mm}\vspace{2mm}

The compression method of loading and unloading is based on the fact that the displacements recovered during unloading are largely elastic, in which case the Young's modulus, $E$, can be simply analyzed from the unloading curve where the Hooke's law holds \cite{Oliver_Pharr1992}. Young's modulus of an elastic material is a ratio of uniaxial stress ($\sigma$) over strain ($\epsilon$) which can be related to an external force $F$ over the uniaxial length changing $\Delta L$ by
\begin{equation}
E=\frac{\epsilon}{\sigma}=\frac{F/A_0}{\Delta L/L_0}=\frac{F}{\Delta L}\cdot\frac{L_0}{A_0}
\end{equation}
where $L_0$ is the initial length and $A_0$ is the initial cross-sectional area of the material without any applied force. 

In this study, taking the initial unloading process into estimation, the Young's modulus can be given by
\begin{equation}
E=\frac{dF}{dh}\cdot\frac{L_0-\Delta L}{A_0+\Delta A}=S\cdot\frac{L_0-h_f}{(w+\nu h_f)^2}
\end{equation}
where $(L_0-h_f)$ is the material length at initial unloading process while $(w+\mu h_f)$ is the material width at initial unloading process with Poisson's ratio $\nu$ (ranged between 0.47 to 0.5 for biomaterials ~\cite{Fung1981, Chen1996, Toyras2001}). The elastic unloading stiffness, $S=dF/dh$, is defined as the slope of the unloading curve during the initial stages of unloading.

A schematic illustration of force-displacement curve is shown in figure 1 where one cycle of loading and unloading of an indenter on a material (which is relatively soft to the indenter) is presented ~\cite{Oliver_Pharr1992}.
\begin{figure}
	\centering % figure 1
 \includegraphics[]{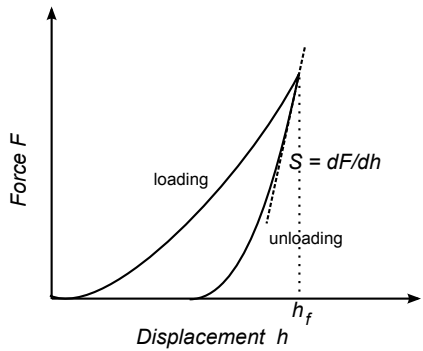} 
	\caption[F-D curve]{Schematic illustration of a cycle of loading and unloading force-displacement curve. The unloading stiffness $S$ is defined as the slope of the force and displacement at the initial unloading process. $h_f$ is the final displacement of the material ~\cite{Oliver_Pharr1992}.}
	\label{fig:FDCurve}
\end{figure}
 
\paragraph{Dynamic Force Moduli}\hspace{18mm}\vspace{2mm}

The stress-strain relationship for a linear viscoelastic material under a sinusoidal loading can be expressed as ~\cite{Oliver_Pharr2008, Oliver_Pharr2004} 
\begin{equation}
\sigma_0cos\omega t=\epsilon_0E'cos\omega t+E''sin(\omega t+\phi)
\end{equation}
where $\sigma_0$ is the stress amplitude, $\epsilon_0$ is the strain amplitude, $\omega$ is the angular frequency, $\phi$ is the phase between the stress and strain. The storage modulus, $E'$, which corresponds to the in-phase term, can be given by [Herbert2008, RA Jones]
\begin{equation}
E'=\frac{\sigma_0}{\epsilon_0}cos\omega t
\end{equation}
and the loss modulus, $E''$, the out-of-phase term, is given by
\begin{equation}
E''=\frac{\sigma_0}{\epsilon_0}sin\omega t.
\end{equation}
The stress and strain can also be expressed in a complex form, i.e., 
\begin{equation}
\sigma*=\sigma_0e^{i\omega t}\\
\epsilon*=\epsilon_0e^{i\omega t+\phi}.
\end{equation}
The complex form of the modulus $E*$ can thus related to the $E'$ and $E''$ by
\begin{equation}
E*=\frac{\sigma}{\epsilon}=\frac{\sigma_0}{\epsilon_0}e^{i\phi}
  = \frac{\sigma}{\epsilon}(cos\phi+sin\phi)= E'+iE''.
\end{equation}

\section{System}
\begin{figure}
	\centering % figure 2
 \includegraphics[]{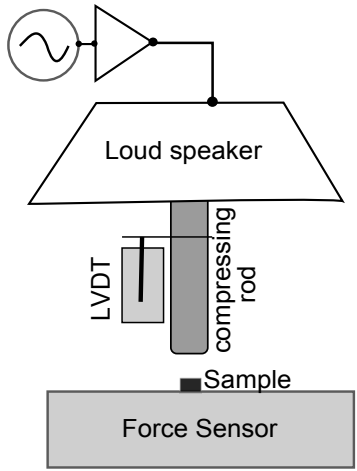} 
	\caption[F-D curve]{The system diagram for measuring the elastic modulus. The compressing rod with flat end is driven by a dynamic loud speaker. The distance is measured by a LVDT (linear variable distance transducer). The sample is placed on top of the force sensor which monitors the force applied on the sample.}
	\label{fig:system}
\end{figure}

The system developed for measuring the elastic properties of small biomaterials is shown in figure 2. A loud speaker (4\", BG\! 10, 8\! $\Omega$ , Visaton, Germany) is used as an actuator for driving the compressing rod. The compressing rod is made of stainless steel with a flat end of diameter 1.25\! cm , and fixed on the diaphram of the speaker. The loud speaker is powered by a modified stereo amplifier (K4003, Velleman-Kit, Beigium) and the input signal is controlled by a computer with LabView programmes and an interfacing DAQ card . 
A liner variable distance transducer (LVDT, Position sensor M-0.5, Applied Measurements Ltd., UK) is used for measuring the displacement of the compressing rod which covers a linear distance of $\pm$\! 0.5\! mm. The digital noise from the LVDT is approximately 4.05 mV which corresponds to 5.05\! $\mu$m. A force sensor modified from a balance (full-scale range 50g, resolution 0.01g, APS-50, Farnell, UK) is placed under the compressing rod. The digital noise from the force sensor is approximately 4.02\! mV which corresponds to 0.09\! mN. The sample is placed on the force sensor and under the compressing rod.

Two different types of silicone rubber (2-mm thick, 4\! mm\! $\times$\! 4\! mm cross-sectional area) have been tested with the system which have Young's Modulus of 0.32\! MPa (794N, DowCorning, USA) and 1.80\! MPa (Bond Flex 100HMA, Bostik, UK). The system gives values within 3\! \% of the published values from the silicone sample datasheets at 21\! $^o$C.

\section{Experiment Setup}

The human bronchial airway tissues were obtained from the surgery-removed lung specimens donated by patients with lung cancer and COPD. All the experiments in this study were under the human tissue act policy in King's College London and approved by the ethical committee in Guy's and St Thomas' NHS hospital. Samples including bronchial airway walls, cartilages, and mucosa were dissected from the tissues and preserved in PBS solution at room temperature (21\! $\pm$\! 1\! $^o$C). Measurements were performed within 4 hours after the surgery at room temperature. The dissected specimens were cut into roughly square shape with a cross sectional area of about 2\! mm\! $\times$\! 2\! mm and placed under the compressing rod with flat end (1.25\! cm diameter) and the sample surface can be fully in contact with the rod during the measurements. 

Young's modulus is estimated from the force-distance curves which the measurements were taken under a constant speed of 0.1\! Hz. The siffness of the instrument was calibrated by testing without any sample before each measurement and the displacement from the instrument was subtracted from the force-distance curve to give the net displacement of the sample. 

The dynamic force measurements were taken under low frequencies (3-45 Hz). The compressing rod was loaded on the sample with knowing force and displacement before performing the frequency sweep to make sure the sample was fully incontact with the rod. The input sinusoidal signal with 5-mV amplitude was powered by the computer corresponging to 4.51\! mN force. The phase between the force and displacement signals was recorded for each frequency with a LabView programme under 1 \! kHz sampling rate. Calibrations on frequency dependent stiffness of the instrument has also been done with frequency sweep at certain load and subtracted from the dynamic force measurements.   

\section{Results}
\begin{figure}
	\centering % figure 3
 \includegraphics[width=\textwidth]{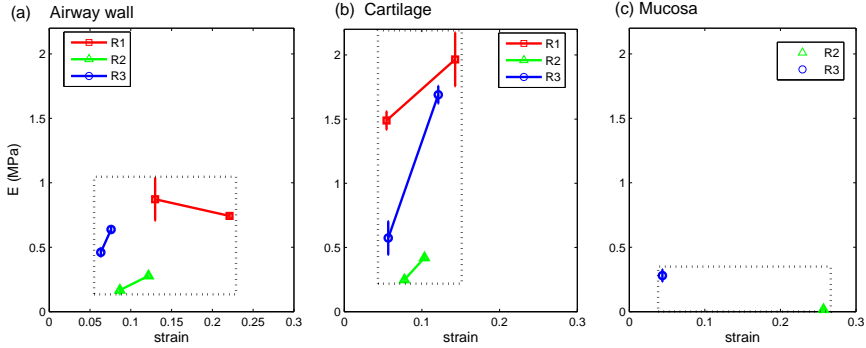} 
	\caption[Young's Modulus]{The Young's Modulus of (a) bronchial airway wall, (b) cartilage and (c) mucosa from three cancer patients R1-3 are plotted against strain.}
	\label{fig:Young's}
\end{figure}
  
\begin{figure}
	\centering % figure 4
 \includegraphics[width=\textwidth]{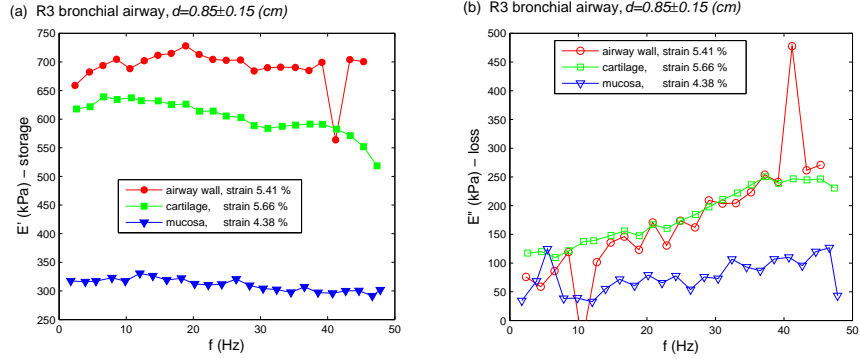} 
	\caption[DMA-R3]{The dynamic force modulus E' and E'' of bronchiole tissue (with diameter 0.85\! $\pm$\! 0.15\! cm) from one cancer patient. The frequency-dependent modulus were compared between the bronchial airway wall, the cartilage, and the mucosa at similar preloaded strains of 5\! $\pm$\! 0.6\! \%.}
	\label{fig:DMA-R3}
\end{figure}

\begin{figure}
	\centering % figure 5
 \includegraphics[width=\textwidth]{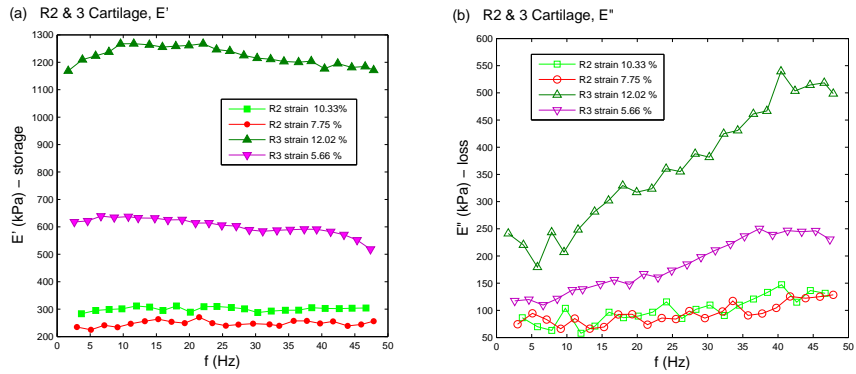} 
	\caption[DMA-Cartilage]{The dynamic force modulus E' and E'' of bronchial cartilage samples from two cancer patients (R2 and R3). The frequency-dependent modulus were compared at different preloaded strains with the same oscillation amplitude.}
	\label{fig:DMA-C}
\end{figure}

Young's modulus with different strain rates of the bronchial airway tissues from three cancer patients are shown in figure 3. Results show the Young's modulus of the airway wall ranges between 0.1 and 1.0\! MPa (with strain from 0.05 to 0.22), and of cartilage ranges between 0.2 and 2.2\! MPa (with strain from 0.05 to 0.15), whereas mucosa of the smallest values ranged betwen 0.35 and 0.01 MPa (with strain between 0.05 to 0.26).
The dynamic force mudulus E' and E'' of bronchial airway wall, cartilage, and mucosa from one patient is plotted in figure 4 with similar preloaded strain. The storage modulus of the airway wall and mucosa samples is found increasing with frequency while decreasing in the cartilage sample with the strain of \~ 5$\pm$0.6\%. Figure 5 shows the dynamic moduli E' and E'' of bronchial cartilage samples from two cancer patients with different preloaded strains. The storage modulus is found increasing slightly with the frequency in one patient while decreasing with frequency between 10-45\! Hz in the other patient. 

\begin{acknowledgement}
We thank Ms Juliet King, Dr Paul Cane, Dr Emma McClean and Dr Amanda Murphey, from Guy's and St Thomas' Hospital for patient recruitment and tissue dissecting. We also thank the Biomedical Research Centre in Guy's and St Thomas' NHS Trust, London, UK for funding this project.  
\end{acknowledgement}

%\appendix

%\section{The First Appendix}

\end{document}